\documentclass[journal = ancac3, manuscript = letter]{achemso}
\usepackage{graphicx,epsf,epsfig}
\usepackage{dcolumn}
\usepackage{bm}
\usepackage{color}
\usepackage{ulem}
\usepackage{subfigure}
\usepackage{upgreek}

\title[\texttt{achemso}]
{Size Dependence of the Multiple Exciton Generation Rate in CdSe Quantum Dots}

\author{Zhibin Lin}
\email{zlin@mines.edu}
\affiliation{Department of Physics, Colorado School of Mines, Golden, CO 80401, USA}
\alsoaffiliation{National Renewable Energy Laboratory, Golden, Colorado 80401, USA}
\author{Alberto Franceschetti}
\email{Alberto.Franceschetti@nrel.gov}
\affiliation{National Renewable Energy Laboratory, Golden, Colorado 80401, USA}
\author{Mark T. Lusk}
\email{mlusk@mines.edu}
\affiliation{Department of Physics, Colorado School of Mines, Golden, CO 80401, USA}

\begin{document}

\begin{abstract}
The multiplication rates of hot carriers in CdSe quantum dots are quantified using an atomistic pseudopotential approach and first order perturbation theory.  Both excited holes and electrons are considered, and electron-hole Coulomb interactions are accounted for.  We find that holes have much higher multiplication rates than electrons with the same excess energy due to the larger density of final states (positive trions). When electron-hole pairs are generated by photon absorption, however, the net carrier multiplication rate is dominated by photogenerated electrons, because they have on average much higher excess energy. We also find, contrary to earlier studies, that the effective Coulomb coupling governing carrier multiplication is energy dependent. We show that smaller dots result in a decrease in the carrier multiplication rate for a given absolute photon energy. However, if the photon energy is scaled by the volume dependent optical gap, then smaller dots exhibit an enhancement in carrier multiplication for a given relative energy.  
\end{abstract}

\maketitle
\newpage

%%%%%%%%%%%%%%%%%%%%%%%%%%%%%%%%%%%%%%%%%%%%%%%%%%%%%%%%%%%%%%%
%Introduction
%%%%%%%%%%%%%%%%%%%%%%%%%%%%%%%%%%%%%%%%%%%%%%%%%%%%%%%%%%%%%%%
Solar cells normally produce a single electron-hole pair per photon absorbed. The excess photon energy $\hbar \omega - E_{gap}$ (where $\omega$ is the photon frequency and $E_{gap}$ is the semiconductor band gap) is converted to heat, and represents a net loss for the photo-conversion efficiency. It is theoretically possible, however, for a photon of sufficiently high energy to generate two or more lower energy excitons, in a process known as multiple exciton generation (MEG). Provided that this process occurs faster than competing carrier relaxation processes, more than one pair of charge carriers might be collected per photon absorbed. This raises the prospect of designing solar-cell devices that exploit MEG in order to utilize high energy photons with greater efficiency than is currently possible.  

Early work in this area~\cite{Nozik_ARPC_2001, Nozik_PE_2002} suggested that the MEG efficiency is enhanced by quantum confinement of carriers, motivating experimental investigations of MEG in different nanostructured materials \cite{Schaller_PRL_2004, Ellingson_NL_2005, Schaller_APL_2005, Schaller_NP_2005, Schaller_NL_2006, Schaller_PRL_2006, Murphy_JACS_2006, Schaller_JPC_2006, Beard_NL_2007, Schaller_NL_2007, Luther_NL_2007, Nair_PRB_2007, Nair_PRB_2008, Trinh_NL_2008, Mcguire_ACR_2008, Ben-Lulu_NL_2008, Beard_NL_2009, Pijpers_NP_2009, Ji_NL_2009, Mcguire_NL_2010, Khafizov_NL_2010}.  Initial reports of very high MEG yields in PbSe, PbS, and CdSe nanocrystals \cite {Schaller_PRL_2004, Ellingson_NL_2005, Schaller_NL_2006, Schaller_APL_2005} were followed by a critical analysis of the assumptions used to extract the MEG yield from optical spectroscopy measurements \cite {AF_PRL_2008, Trinh_NL_2008, Mcguire_ACR_2008, Mcguire_NL_2010}. The consequences of the energy dependence of the absorption cross section \cite{Trinh_NL_2008}, the non-linear dependence of the bleaching ratio on the number of excitons \cite{AF_PRL_2008}, and the non-radiative recombination of charged excitons \cite{Mcguire_ACR_2008} on the accuracy of MEG measurements have all been discussed in the literature. This has led to a re-assessment of earlier experimental results.    

A controversy has recently emerged, though, as to how much, if at all, quantum confinement really enhances MEG. Most experimental investigations of MEG in nanostructures have focused on the MEG quantum yield (QY), defined as the average number of excitons produced per absorbed photon. MEG manifests itself by producing more than one exciton (QY>1) when the photon energy exceeds a certain threshold, $\hbar \omega_{th}$. Experiments for PbSe \cite{Schaller_PRL_2004} and Si \cite {Beard_NL_2007} nanocrystals have revealed that the QY measured at a given {\it absolute} photon energy decreases when the nanocrystal size decreases, while other experiments have reported a weak and/or non-monotonic dependence of the QY on size \cite{Nair_PRB_2008}. When measured with respect to the {\it relative} photon energy, $\hbar \omega / E_{gap}$, however, the QY increases as the nanocrystal size decreases \cite {Schaller_PRL_2004, Ellingson_NL_2005, Beard_NL_2007, Schaller_NL_2007, Mcguire_ACR_2008}. Rabani et al. \cite {Rabani_NL_2008} calculated the QY of a set of small CdSe and InAs  nanocrystals and found an increase as the size decreases. A few experimental \cite{Beard_NL_2007, Nair_PRB_2008, Pijpers_NP_2009} and theoretical \cite{Pijpers_NP_2009, Delerue_PRB_2010} investigations have also attempted a comparison of the QY of quantum dots vs. bulk semiconductors. Beard et al. \cite{Beard_NL_2007} reported a significantly higher QY in Si nanocrystals than in bulk Si at the same relative photon energy. Nair et al. \cite {Nair_PRB_2008} found that the QY of PbS nanocrystals is lower than that of PbS cystalline films \cite {Smith_JOSA_1958} when measured at fixed absolute photon energy. Similar results where reported by Pijpers et al. \cite {Pijpers_NP_2009}, Delerue et al. \cite {Delerue_PRB_2010} and McGuire et al. \cite {Mcguire_NL_2010} for PbS and PbSe nanocrystals. Notwithstanding these results, systematic experimental or theoretical studies of the size dependence of MEG are not available. Furthermore, the QY depends on to the ratio of the MEG rate, $\Gamma_{M\!E\!G}$, and the total exciton decay rate, $\Gamma_{tot}$. The latter includes competing decay processes, such as phonon-assisted carrier cooling. Since both $\Gamma_{M\!E\!G}$ and $\Gamma_{tot}$ depend on the nanostructure size as well as the photon energy, the observation of a certain dependence of the QY on  size does not necessarily imply that the same trend exists for $\Gamma_{M\!E\!G}$. Thus, it would be very useful to quantify the size dependence of the MEG {\it rate} itself. This quantity is difficult to measure experimentally, but accurate calculations of $\Gamma_{M\!E\!G}$ as a function of size can provide information on multi-exciton dynamics and clarify the role of quantum confinement in the MEG process.  

%%%%%%%%%%%%%%%%%%%%%%%%%%%%%%%%%%%%%%%%%%%%%%%%%%%
% Results
%%%%%%%%%%%%%%%%%%%%%%%%%%%%%%%%%%%%%%%%%%%%%%%%%%%
\subsection {Results and Discussion} 

We use atomistic pseudopotential calculations, coupled with first-order perturbation theory, to quantify the dependence of the MEG rate on photon energy and  nanoparticle volume for zinc-blende CdSe nanocrystal quantum dots (QDs). The volume dependence of the MEG rate is decomposed to show the relative contributions of Coulomb matrix elements and final density of states. Both hole and electron contributions to the MEG rate of an electron-hole pair are accounted for, so that we can make clear which of these dominates MEG. We find that while the hole MEG rate is significantly larger than the electron MEG rate for a given excess energy, the electron decay dominates the MEG of photo-generated electron-hole pairs because of the larger excess energy of the electron. Tables 1-3 describe the energy and size dependence of the electron, hole, and exciton MEG rates of CdSe QDs. We find that (i) the MEG rate increases rapidly with the excess energy of the carriers. (ii) At fixed {\it absolute} photon energy, the MEG rate of a photo-generated electron-hole pair decreases as the volume of the QD decreases. However (iii) at fixed {\it relative} photon energy, the MEG rate increases as the volume decreases.

%%%%%%%%%%%Figure%%%%%%%%%%
\begin{figure}
\centering
 \includegraphics[width=.6\textwidth, keepaspectratio, clip]{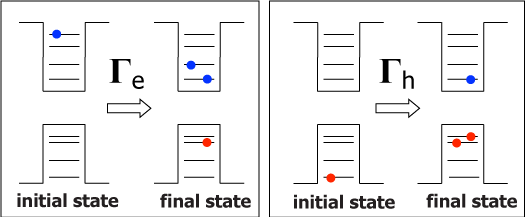}
\begin{center}  
\caption{Schematic diagram of MEG decay paths for an excited electron (left) and an excited hole (right). In practice, all possible combinations of electrons and holes in the final trion states are considered in the calculation of MEG rate for a given initial, excited state.}
\label{fig:CM2QDS}
 \end{center}
\end{figure}
%%%%%%%%%%end figure%%%%%%%%%%%%%%%

When a carrier has sufficient excess energy (larger than $\sim E_{gap}$) it can spontaneously decay into a trion in a process known as {\it impact ionization} (see Fig. 1). The impact ionization rates for electrons and holes are calculated here by applying Fermi's golden rule to the formation of trion states. We consider here four nearly spherical QDs: Cd$_{216}$Se$_{213}$ (diameter D=2.9 nm), Cd$_{312}$Se$_{321}$ (D=3.2 nm), Cd$_{484}$Se$_{495}$ (D=3.7 nm), and Cd$_{784}$Se$_{739}$ (D=4.3 nm). Figure 2 shows the calculated MEG rates of electrons and holes for the two CdSe QDs considered here (diameter 2.9 and 3.7 nm, respectively). The MEG rates are plotted in Fig. 2 as a function of the absolute electron/hole energy measured with respect to the vacuum level (Figs. 2a and 2b), as well as the relative electron/hole excess energy $|E - E_{0}|/E_{gap}$ (Figs. 2c and 2d). Here $E_0$ is the conduction-band minimum (CBM) in the case of electrons and the valence-band maximum (VBM) in the case of holes, and  the optical gap $E_{gap}$ is calculated as $E_{gap}=\varepsilon_{CBM} - \varepsilon_{VBM} - J_{VBM,CBM}$.  Fig. 2 shows that the MEG rate of electron and holes increases rapidly as their excess energy increases. Fig. 2 also shows that, for the same relative excess energy, excited holes have transition rates more than an order of magnitude greater than excited electrons. 

%%%%%%%%%%%figure%%%%%%%%%
\begin{figure}
\centering
\includegraphics[width=.75\textwidth]{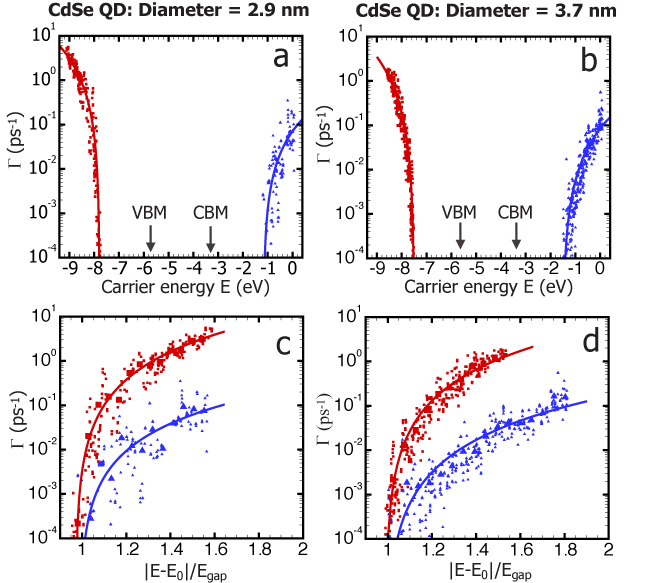} 
\begin{center}
\caption{Calculated MEG rate as a function of the electron and hole energy for CdSe QDs with diameters of 2.9 nm (left) and 3.7 nm (right). Energies are shown in two ways. (a) and (b): with respect to the vacuum level; (c) and (d): with respect to the band-edge energy $E_0$ (CBM for electrons and VBM for holes) and rescaled by the optical band gap $E_{gap}$. The small symbols correspond to the transition rate of individual electron (blue) or hole (red) initial states to final trion states, while the large symbols in the bottom figures represent the arithmetic average of initial states within a 100 meV energy window. Solid lines, $\Gamma(E)$, are fitted using the averaged data (large symbols). Arrows indicate the positions of the CBM and VBM.}
\label{fig2}
\end{center}\end{figure}
%%%%%%%%%%end figure%%%%%%%%%%%%%%%

The calculated MEG rates for all the QD sizes considered here are summarized in Fig. 3a. For each QD, the energy dependence of the electron and hole MEG rates were fitted to a power law, $\Gamma_{h,e} (E) \propto (E-E_{th})^{\alpha_{h,e}}$, where the threshold energy $E_{th}$ is the ground-state energy of a negative trion (for electrons) or a positive trion (for holes), calculated according to \ref{eq:Ef_e} and \ref{eq:Ef_h}, respectively. Table 1 summarizes the values of $\alpha_h$ and $\alpha_e$ extracted from the fit. We find that $\alpha$ varies between 2.1 and 2.7, depending on the QD size, and is slightly larger for holes than for electrons. The volume dependence of the MEG rates was calculated for a set of given electron and hole energies and fitted to the function $\Gamma_{h,e} (V) = \Gamma^0_{h,e} + B_{h,e} V^{\beta_{h,e}} $, where $\Gamma^0_{h,e}$, $B_{h,e}$ and $\beta_{h,e}$ are fitting parameters listed in Table 2.  In particular, $\beta_{h,e}$ ranges between -2.0 and -2.4 for holes (depending on their energy) and between -1.2 and -1.8 for electrons.

 \begin{table}
 \caption{Fitted parameters for the energy dependence of electron and hole MEG rates: $\Gamma_{h,e} (E) = A_{h,e} (E-E_{th})^{\alpha_{h,e}}$ as shown in Figure 3a.}
 \label{tab1}
 \begin{tabular}{|c|c|c|}\hline
 Diameter (nm)&$\alpha_{h}$&$\alpha_{e}$\\ \hline
2.9 &2.3 & 2.1 \\
3.2 &2.4 & 2.3 \\
3.7 &2.7 & 2.5 \\
4.3 &      & 2.6 \\ \hline
 \end{tabular}
 \end{table}
 
  \begin{table}
 \caption{Fitted parameters for the volume dependence of electron and hole MEG rates: $\Gamma_{h,e} (V) = \Gamma^0_{h,e} + B_{h,e} V^{\beta_{h,e}} $ at given carrier energy, $E$.}
 \label{tab2}
 \begin{tabular}{|ccc|ccc|}\hline
 $E (eV)$&$\beta_{h}$&$B_{h}$&$E (eV)$&$\beta_{e}$&$B_{e}$\\ \hline
-7.8 & -2.0 & -11.2 & -0.6 & -1.8 & -0.93\\
-8.0 & -2.4 & -66.8 & -0.7 & -1.7 & -0.78\\
-8.2 & -2.2 & -69.7 & -0.8 & -1.6 & -0.60\\
-8.4 & -2.3 & -128.0 & -0.9 & -1.2 & -0.18\\ \hline
 \end{tabular}
 \end{table}

In order to explain the trends displayed in Fig. 3a, we decompose the MEG rate into the trion density of states (DOS) contribution, $\rho(E)=\sum_{f}\delta(E-E_{f})$, and the effective Coulomb coupling contribution, $\vert \bar W \vert ^{2} =\hbar\Gamma(E)/2\pi \rho(E)$. These two components are shown in Figs. 3b and 3c, respectively. The trion DOS (Fig. 3b) shows an energy dependence very similar to the overall MEG rate of Fig 3a, i.e., the number of final trion states increases rapidly as the excess energy increases. While this trend holds for trions deriving from both holes and electrons, the positive trion DOS is significantly larger than the negative trion DOS, as shown in the inset plot of Fig. 3b for a 3.7 nm dot. The reason for this behavior is that the manifold of hole states is much denser than its electron counterpart. As a result, the number of available trion states, for a given excess energy,  is  much higher for holes than electrons, and this dominates the overall MEG rates. Previous calculations using an atomic pseudopotential approach for PbSe QDs have revealed a dense feature of valence states \cite{Franceschetti_NL_2006}. In the case of CdSe QD, the asymmetry in the density of states between the conduction and valence states is even more pronounced. 
%%%%%%%%%%%figure%%%%%%%%%
\begin{figure}[ptb]\begin{center}$
\begin{array}{cc}
\includegraphics[width=0.4\textwidth]{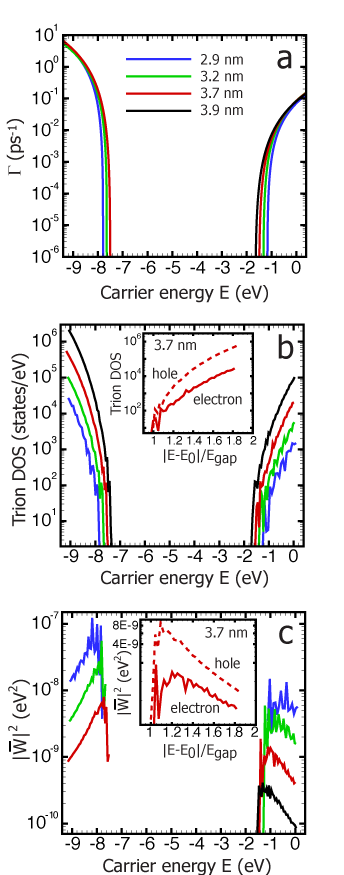}
\end{array}$
\end{center}
\begin{center}
\caption{(a) MEG rate $\Gamma(E)$, (b) trion density of states $\rho_{trion}(E)$, and (c) effective Coulomb coupling, $\vert \bar W \vert ^{2} (E)$, as a function of the electron and hole energy for four CdSe QDs. Insets in (b) and (c) show the trion DOS and the effective Coulomb coupling for a 3.7 nm QD, as a function of the relative energy rescaled by the optical band gap. The MEG rates are fitted using averaged data in the manner described in Figure 2. The trion DOS is obtained using a Gaussian broadening of 10 meV. For the largest dot (4.3 nm), the MEG rate and effective Coulomb coupling are only shown for electrons due to the significant computational cost required to calculate MEG rates from the large set of possible positive trion configurations.}
\label{fig3}
\end{center}\end{figure}
%%%%%%%%%%end figure%%%%%%%%%%%%%%%

The effective Coulomb coupling, $|\bar W|^2$, (Fig. 3c) peaks just above the threshold energy and then exhibits a decay as the energy increases. The decay becomes faster as the size of the QD increases.
The size dependence of the effective Coulomb coupling was obtained by fitting the data for all four CdSe dots to a power law: $\vert \bar W \vert ^{2} \propto V^{\gamma}$. The Coulomb coupling scaling is energy dependent with $\gamma$ ranging from -2.6 to -3.2 for electrons and -3.6 to -4.0 for holes. 
This trend holds for all four QDs and is in contrast to the assumptions and conclusions of earlier studies. Specifically, a constant Coulomb coupling was assumed for an examination of impact ionization in bulk Si~\cite{Kane_PR_1967}, and a recent tight-binding study concluded that there is only a weak energy dependence in the Coulomb coupling for PbSe QDs~\cite{Allan_PRB_2008}. Recent pseudopotential calculations for CdSe QDs \cite{Rabani_NL_2008} found that the Coulomb coupling has no energy dependence for dots up to 3.0 nm. Our calculations suggest that this is not the case for larger QDs. In any case, the downward trend in the energy dependence of the Coulomb coupling is overshadowed by the opposing trend in the trion DOS.  

To understand the decrease in the effective Coulomb coupling with increasing excess energy, we analyze the spatial distribution of the initial-state electron and hole wave functions. For each single-particle state $i$ of a QD, we calculate the fraction of the wave function norm that resides in the surface region of the QD, $S_i = \int_{{\bf r} \in S} |\psi_i ({\bf r})|^2 d{\bf r}$, where $S$ is a thin shell around the surface of the QD. Figure 4 shows $S_i$ (Fig. 4a) and the charge density distribution of hole states located between -8.0 eV and -7.5 eV (Fig. 4b) for the 3.7 nm CdSe QD. The plot of $S_i$ (Fig. 4a)  indicates that states near the band edges are localized in the core of the QD. As the energy level moves away from the band edges, though, the wave functions spread to occupy a large portion of the surface shell. The broad peak located 3 eV below the VBM originates from surface states associated with the ligands used to passivate the QD surface (Fig. 4b). Also shown in Fig. 4a is the Coulomb coupling $|\bar W|^2$ of the same 3.7 nm QD (already shown in Fig. 2b). In the case of the electron levels, $|\bar W|^2$ decreases as the surface projection, $S_i$, increases. However, this is not the case for the hole levels which exhibit large values of $|\bar W|^2$ even when their wave functions are largely localized near the surface of the QD (Fig. 4b). Fig. 4 suggests that there is no obvious correlation between $|\bar W|^2$ and the core/surface projection of the wave functions. Surface localization of the wave functions does not appear to have a dramatic effect on the MEG rate. The dependence of the MEG {\it quantum yield} on surface passivation, reported e.g. by Beard et al. \cite {Beard_NL_2009}, appears to be related to the effects of surface passivation on competing decay channels for the photo-excited electron-hole pair.

%%%%%%%%%%%figure%%%%%%%%%
\begin{figure}[h]
\begin{center}$
\begin{array}{cc}
\includegraphics[width=0.4\textwidth]{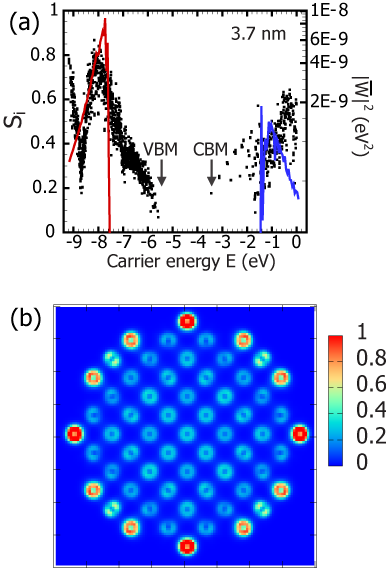} 
\end{array}$
\label{fig:fig4}
\end{center}
\begin{center}
\caption{(a) Surface projection fraction, $S_{i}$, of the single-particle wave function norm plotted as a function of the electron and hole energy. Each point corresponds to a single-particle eigenstate. The surface region is defined as the set of points closer than $R_{WS}$ (the Wigner-Seitz radius of bulk CdSe) from at least one surface Cd/Se atoms. The effective Coulomb coupling, $\vert \bar W \vert ^{2}$, is shown (solid line) for the region where MEG occurs. (b) Charge density distribution (arbitrary units) of hole states between -8.0 eV and -7.5 eV of a 3.7 nm CdSe QD, plotted on a (001) plane. }
  \end{center}
\end{figure}
%%%%%%%%%%end figure%%%%%%%%%%%%%%%

Having discussed the MEG rates of individual electrons and holes, we now consider the MEG rate of a single exciton decaying into a biexciton. Because many-body selection rules forbid the simultaneous transition of the electron and the hole (in first-order perturbation theory), the MEG rate of an electron-hole pair (h, e) is simply the sum of the MEG rates of the electron and the hole:
$\Gamma(h,e) = \Gamma_h + \Gamma_e$, where we have assumed that $\Gamma_h$ ($\Gamma_e$) is weakly perturbed by the presence of a spectator electron (hole). When the QD is excited by a laser pulse, different e-h pairs that are nearly degenerate in energy can be excited. Thus, we calculate the MEG rate for a photon of energy, $\hbar \omega$, as

\begin{equation}
\label{eq:prob1}
\Gamma(\hbar\omega)=\sum_{e}p_{e}(\hbar\omega)\Gamma_{e} + \sum_{h}p_{h}(\hbar\omega)\Gamma_{h} 
\end{equation}
where \begin{equation}
\label{eq:prob2}
p_{h}(\hbar\omega)=\frac{\sum_{c}\vert M_{h,c}\vert ^{2}\delta(E_{h,c}-\hbar\omega)}{\sum_{v,c}\vert M_{v,c}\vert ^{2}\delta(E_{v,c}-\hbar\omega)}
\end{equation}
\begin{equation}
\label{eq:prob3}
p_{e}(\hbar\omega)=\frac{\sum_{v}\vert M_{v,e}\vert ^{2}\delta(E_{v,e}-\hbar\omega)}{\sum_{v,c}\vert M_{v,c}\vert ^{2}\delta(E_{v,c}-\hbar\omega)}
\end{equation}
Here $p_h$ ($p_e$) is the probability of creating a hole in the valence state v (an electron in the conduction state c) as a result of the absorption of a photon of energy $\hbar\omega$.  The integral $M_{v,c}=\langle \psi_{v}\vert \boldsymbol{r} \vert \psi_{c}\rangle$ is the dipole matrix element that characterizes the transition from the valence state v, to the conduction state c, and $E_{v,c} $= $\varepsilon_{c}$ - $\varepsilon_{v}$ - $J_{v,c}$.

%%%%%%%%%%%figure%%%%%%%%%
\begin{figure}[ptb]\begin{center}$
\begin{array}{cc}
\includegraphics[width=.75\textwidth]{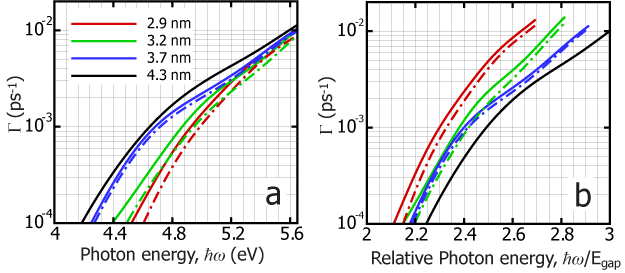} 
\end{array}$
\end{center}
\begin{center}
\caption{MEG rate [\ref{eq:prob1}] as a function of (a) absolute photon energy and (b) relative photon energy scaled by the optical band gaps for four CdSe QDs. Dashed lines are calculated including only the contribution from excited electrons, i.e., $\Gamma_{e}$ in \ref{eq:prob1}. For the largest dot (4.3 nm), the MEG rate includes only the electron contribution, as the hole contribution is negligible (see text).}
  \label{fig5}
\end{center}\end{figure}
%%%%%%%%%%end figure%%%%%%%%%%%%%%%

Figure 5a shows the MEG rate for all four CdSe QDs as a function of the photon energy, $\hbar \omega$, calculated using \ref{eq:prob1}. As expected (see e.g. Fig. 2), the MEG rate increases as the photon energy increases. For a given absolute photon energy, the MEG rate increases as the dot size increases (Fig. 5a). However, the trend is just the opposite if the photon energy is rescaled by the optical band gap of each QD (Fig. 5b). This is the result of quantum confinement which causes the optical gap to increase with decreasing QD size, which implies that the photon energy has to increase to keep the ratio $\hbar \omega / E_{gap}$ constant. The optical band gap of the four CdSe QDs in this work decreases from 2.14 eV (2.9 nm diameter) down to 1.86 eV (4.3 nm diameter). The calculated MEG rate is relatively low, e.g. $\Gamma < 10^{-2}$ ps$^{-1}$ at $\hbar \omega = 2.7 E_{gap}$ for the 3.2 nm QD. This result suggests that, at low photon energies, MEG can be easily overcome by competing carrier relaxation processes, in agreement with the experimental results of Nair {\it et al.} \cite{Nair_PRB_2008}.

Figure 5 also makes clear that the main contribution to the total MEG rate originates from electrons decaying into negative trions. The reason is that, because of the distribution of dipole matrix elements [\ref{eq:prob2} and \ref{eq:prob3}], the excess energy of the photo-generated electron is larger than that of the hole, so the electron has much higher probability of decaying by MEG, despite the fact that the MEG rate for an individual hole is larger than that of an electron with the same excess energy (Fig. 2).  If an excited electron could be efficiently converted to an excited hole through an Auger process, though, the total MEG rate could be significantly enhanced. 

%%%%%%%%%%%figure%%%%%%%%%
\begin{figure}[ptb]\begin{center}$
\begin{array}{cc}
\includegraphics[width=.75\textwidth]{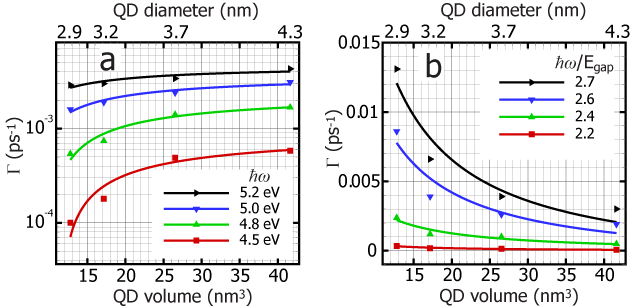} 
\end{array}$
\end{center}
\begin{center}
\caption{Volume dependence of the MEG rate of CdSe QDs, calculated (a) at fixed 
absolute photon energy and (b) at fixed relative photon energy 
scaled by the optical band gap. Solid lines are power law fits to the MEG rate for 
prescribed energies, and the resulting coefficients are collected in Table 3.}
  \label{fig6}
\end{center}\end{figure}
%%%%%%%%%%end figure%%%%%%%%%%%%%%%

The trends in size scaling can be made more precise by fitting the results to power law functions of the QD volume. These fits are given in Fig. 6. We find that, for a given absolute photon energy, the MEG rate scales as  $\Gamma=\Gamma_{0}+\alpha V^{\beta}$ with $\beta$ ranging from -0.70 to -0.91 in the energy range under consideration. In contrast, the MEG rate, as a function of the relative photon energy, scales as $\Gamma=\alpha V^{\beta}$ with $\beta$ lying between -1.36 and -1.21 up to 2.7$E_{gap}$ (Table 3). It has been argued~\cite{Nair_PRB_2008} that the MEG rate should not exhibit a significant volume dependence because of the offsetting volume dependencies of Coulomb coupling and trion density of states. Specifically, it was suggested that the average Coulomb coupling scales as $V^{-3}$ while the trion DOS scales as $V^{3}$. Our  calculations show, however, that the volume dependence of the Coulomb coupling is roughly $V^{-3}$ for excited electrons but closer to $V^{-4}$ for holes. The trion DOS scaling depends on photon energy and is $V^{\alpha}$ with $\alpha$ ~3.1 to 4.1, while for the excited holes $\alpha$ ranges from 3.8 to 5.5. These results show that the volume scaling contributions to MEG rate from the effective Coulomb coupling and trion DOS do not cancel each other and there is an energy dependence that must be considered as well. The net scaling of the MEG rate is quantified in Fig. 6 and Table 3.

 \begin{table}
 \caption{\label{tab3}Fitted parameters for the volume dependence of MEG rate: (left) $\Gamma=\Gamma_{0}+\alpha V^{\beta}$ at given absolute photon energy, $\hbar\omega$; (right) $\Gamma=\alpha V^{\beta}$ at given photon energy scaled by the optical band gap of the QD, $\hbar\omega/E_{gap}$, as shown in Figure 6.}
 \begin{tabular}{|cc|cc|}\hline
 $\hbar\omega (eV)$& $\beta$&$\hbar\omega/E_{gap}$&$\beta$\\ \hline
4.5 & -0.70 & 2.2 & -1.36\\
4.8 & -0.78 & 2.4 & -1.21\\
5.0 & -0.82 & 2.6 & -1.20\\
5.2 & -0.91 & 2.7 & -1.21\\ \hline
 \end{tabular}
 \end{table}
 
While the choice of plotting the MEG rate versus absolute or relative photon energy (see e.g. Figs. 5a and 5b) may appear arbitrary, it has been argued that for a given QY the MEG {\it efficiency} increases as the excitonic gap increases, because the energy of the multi-excitons produced by MEG increases \cite {Delerue_PRB_2010, Mcguire_NL_2010}. Beard et al. \cite {Beard_NL_2010} have suggested that the MEG efficiency is proportional to the slope of the QY vs. the relative photon energy $\hbar \omega / E_{gap}$ (above the MEG threshold $\hbar \omega_{th}$). Thus, the dependence of the MEG rate on QD volume calculated at fixed relative photon energy (Fig.  6b) is more appropriate to describe the trend of the MEG efficiency with QD size than the MEG rate calculated at absolute photon energy (Fig. 6a).

In summary, we have applied an atomistic pseudopotential method to quantify the relation between size and MEG rate in CdSe QDs. The MEG rate increases as QD size decreases only if the comparison is made for the same relative photon energies, i.e., after normalization using the volume dependent optical gap. The opposite trend is found, however, if the comparison is made using the same absolute photon energy for each QD. Both the effective Coulomb coupling and the trion DOS are energy dependent with the latter being a stronger function of energy. Although the decay rate of an excited hole is much larger than that of an excited electron, the total MEG rate is dominated by photo-excited electrons, because they have higher excess energy than photo-excited holes. Volume scaling laws for overall MEG, trion DOS, and Coulomb coupling integral have now been quantified under realistic conditions and with a minimal number of simplifying idealizations.

\subsection{Methods}
Our computational approach utilizes an atomistic pseudopotential that is tailored to capture the experimental electronic structure of CdSe QDs~\cite{Wang_PRB_1996, Wang_JPCB_1998, Franceschetti_PRB_1999}. The CdSe quantum dots are constructed by carving out a Se-centered sphere from bulk CdSe zinc-blende structure. The bulk crystal utilizes the experimental lattice constant of $a_0$=6.08 \AA \cite{Kaxiras_2003}.  All surface Cd and Se atoms are passivated with ligand-like atomic pseudopotentials that are fitted to remove the surface states from the band gap of the QD. The Cd and Se atomic pseudopotentials are fitted to experimental bulk CdSe properties such as transition energies and effective masses and to single-particle bulk wavefunctions calculated from first principles \cite {Wang_PRB_1996, Wang_JPCB_1998}. 
The single-particle energies, $\varepsilon_i$, and wave functions, $\psi_i$, are then obtained by solving the single-particle Schr\"odinger equation:
\begin{equation}
[-\frac{\hbar}{2m}\nabla^{2}+V(\boldsymbol{r})+\hat{V}_{SO}]\psi_{i}(\boldsymbol{r}, \sigma)=\epsilon_{i}\psi_{i}(\boldsymbol{r}, \sigma) 
\end{equation}
Here the single particle wave functions,  $\psi_i$, are expanded in a plane wave basis set, $\hat{V}_{SO}$ is the nonlocal spin-orbit operator, and the local potential, $V(\boldsymbol{r})$, is a linear superposition of screened pseudopotentials centered at the atomic positions, \{$\boldsymbol{R}$\}, of atomic type $\alpha$:
\begin{equation}
V(\boldsymbol{r})=\sum_{\alpha,\boldsymbol{R}}\upsilon_{\alpha}(\boldsymbol{r}-\boldsymbol{R})
\end{equation}
The impact ionization rates for electrons and holes are calculated here by applying Fermi's golden rule to the formation of trion states: 
\begin{equation}
 \label{eq:FGR1}
\Gamma_{e} = \frac{2\pi}{\hbar}\sum_{f}\vert\langle e\vert W\vert f\rangle\vert^{2}\delta(E^-_{f}-E_{e})
\end{equation}
\begin{equation}
 \label{eq:FGR2}
\Gamma_{h} = \frac{2\pi}{\hbar}\sum_{f}\vert\langle h\vert W\vert f\rangle\vert^{2}\delta(E^+_{f}-E_{h})
\end{equation}
Here $\vert e\rangle$ and $\vert h\rangle$ are the initial, excited states for an electron in the conduction band and a hole in the valence band, respectively.  The operator, W, is the screened Coulomb matrix element between the initial and final states \cite{Kane_PR_1967,Landsberg_1991,Franceschetti_PRL_2003}. The final trion states are $\vert f\rangle = \vert e_{1}, e_{2}, h_{1}\rangle$ and $\vert f\rangle = \vert e_{1}, h_{1}, h_{2}\rangle$, respectively, and their energies are calculated as 
\begin{equation}
\label{eq:Ef_e}
E^-_f = (\varepsilon_{e1}+\varepsilon_{e2}-\varepsilon_{h1}) + (J_{e1,e2}-J_{e1,h1}-J_{e2,h1}),
\end{equation}
and
\begin{equation}
\label{eq:Ef_h}
E^+_f = (\varepsilon_{e1}-\varepsilon_{h1}-\varepsilon_{h2}) + (J_{h1,h2}-J_{e1,h1}-J_{e1,h2}).
\end{equation}
The screened Coulomb interactions, $J_{i,j}$ are given by
\begin{equation}
\label{eq:Coulomb}
J_{i,j} = \sum_{\sigma,\sigma\prime}\int\int {\rm d}{\bf r} {\rm d} {\bf r}' \ \vert \psi_{i}({\bf r}',{\bf \sigma}')\vert^2 \frac{e^2}{\epsilon({\bf r},{\bf r}')\vert {\bf r} - {\bf r}'\vert }\vert \psi_{j}({\bf r},{\bf \sigma})\vert^2 ,
\end{equation}
where $\epsilon({\bf r},{\bf r}')$ is the dielectric screening function of the QD. \cite{Franceschetti_PRB_1999} We adopt a Lorentzian line shape, $\delta(E_{f}-E_{i}) \rightarrow \frac{1}{\pi}\frac{(\gamma/2)}{(E_{i}-E_{f})^2+(\gamma/2)^2}$, in evaluating \ref{eq:FGR1} and \ref{eq:FGR2}, to account for phonon broadening effects. A value of $\gamma  = 10$ meV is used for the summation over the final states\cite{Franceschetti_PRL_2003}; changes in $\gamma$ from 5 meV to 20 meV were found to not meaningfully affect MEG rates.
Rabani et al. \cite{Rabani_NL_2008} used a pseudopotential approach to calculate the MEG rates of small (up to ~500 atoms) CdSe and InAs nanocrystals. Our approach builds on their method by (i) including spin-orbit coupling, which is significant in CdSe quantum dots, and (ii) including excitonic interactions in \ref{eq:FGR1} and \ref{eq:FGR2}. Electron-hole Coulomb interactions [\ref{eq:Coulomb}] lower the energy of the final (trion) states [see \ref{eq:Ef_e} and \ref{eq:Ef_h}], thereby increasing the density of final states which are in resonance with the initial state. The net result is a significant increase in the MEG rate (an order of magnitude) compared to the case where Coulomb interactions are neglected.

\begin{acknowledgement}
We are grateful to A. Nozik and M. Beard for useful discussions 
concerning MEG efficiency. This work was supported by the Renewable 
Energy Materials Research Science and Engineering Center (NSF Grant No. 
DMR-0820518) at the Colorado School of Mines and the National Renewable 
Energy Laboratory. The calculations were carried out using the high 
performance computing resources provided by the Golden Energy Computing 
Organization at the Colorado School of Mines (NSF Grant No. CNS-0722415).
\end{acknowledgement}

\bibliographystyle{achemso}
%\bibliography{MEG}

\providecommand{\refin}[1]{\\ \textbf{Referenced in:} #1}

\end{document}